\documentclass[final,3p,times]{elsarticle}

\usepackage{amssymb}
\usepackage{amsthm}
\usepackage{lineno}
\biboptions{sort&compress}
\usepackage{hyperref}

\journal{Physics Letters B}

\begin{document}

\begin{frontmatter}

\title{Comparisons of electric charge and axial charge meson cloud distributions
in the PCQM}

\author[SUT,BHU]{X. Y. Liu\corref{cor}}
\ead{lxy\_gzu2005@126.com}

\author[SUT]{K. Khosonthongkee}
\author[SUT]{A. Limphirat}
\author[SUT]{Y. Yan}

\cortext[cor]{Corresponding author}

\address[BHU]{School of Mathematics and Physics, Bohai University, Liaoning 121013, China}
\address[SUT]{School of Physics and Center of Excellence in High Energy Physics and Astrophysics, Suranaree University of Technology, Nakhon Ratchasima 30000, Thailand}

\begin{abstract}
The meson cloud distributions in $r$-space are extracted from the nucleon electromagnetic and axial form factors which are derived in the perturbative chiral quark model. The theoretical results indicate that the electric charge and axial charge distributions of the three-quark core are the same, the magnetic charge distributions of the meson cloud and three-quark core are more or less in the same region and peak at distances of around 2 $\rm GeV^{-1}$, but the axial charge meson cloud distributes mainly inside the three-quark core.
\end{abstract}


\end{frontmatter}

%\linenumbers

\section{\label{sec:intro}Introduction}

The meson cloud of the nucleon, undoubtedly, plays a relevant role in the study of low energy electroweak properties of the nucleon. The meson cloud model, where the nucleon is considered as a system of three valence quarks surrounded by a meson cloud \cite{Theberge:1980,Thomas:1981,Chin:1982,Oset:1984,Gutsche:1989,Dziembowski:1997,Speth:2002,Faessler:2006,Julia:2006,Chen:2007}, has recently been employed to study the generalized parton distribution \cite{Pasquini:2006,Pasquini1:2007}, nucleon electroweak form factors \cite{Lyubovitskij:2001,Lyubovitskij1:2001,Khosonthongkee:2004,Pasquini2:2007,Ramalho:2011,Ramalho1:2013}, nucleon strangeness \cite{Carvalho:2005,Chen:2010}, etc. In Refs.~\cite{Lu:1998,Glozman:1999,Rinehimer:2009}, meson cloud contributions to the neutron charge form factor have been studied and discussed in the meson cloud model, while the effects of the meson cloud on electromagnetic transitions have been estimated in Refs.~\cite{Ramalho:2012,Ramalho:2013,Ramalho:2014}. In our previous works~\cite{Liu:2014,Liu:2015}, the electromagnetic and axial form factors as well as electroweak properties of octet baryons have been studied in the perturbative chiral quark model (PCQM). The theoretical results in the PCQM with the predetermined quark wave functions are in good agreement with the experimental data and lattice QCD values. In addition, Ref.~\cite{Liu:2015} reveals that the meson cloud plays an important role in the axial charge of octet baryons, contributing 30\%--40\% to the total values, and the similar effects have been also observed in other frameworks~\cite{Franklin:2002,Ramalho:2016}.

\begin{figure}[t!]
\begin{center}
\includegraphics[width=0.47\textwidth]{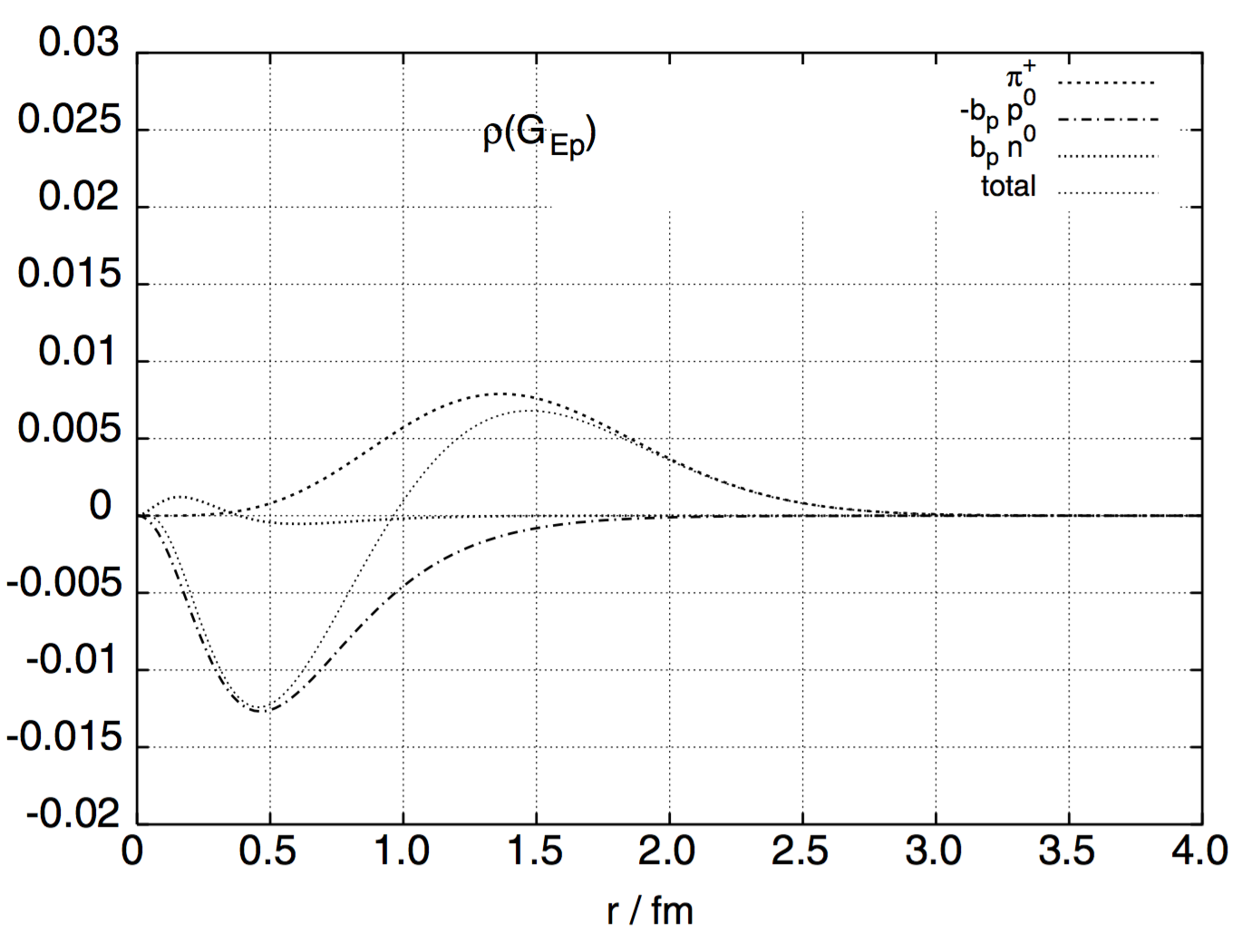}
\hspace{0.5cm}
\includegraphics[width=0.476\textwidth]{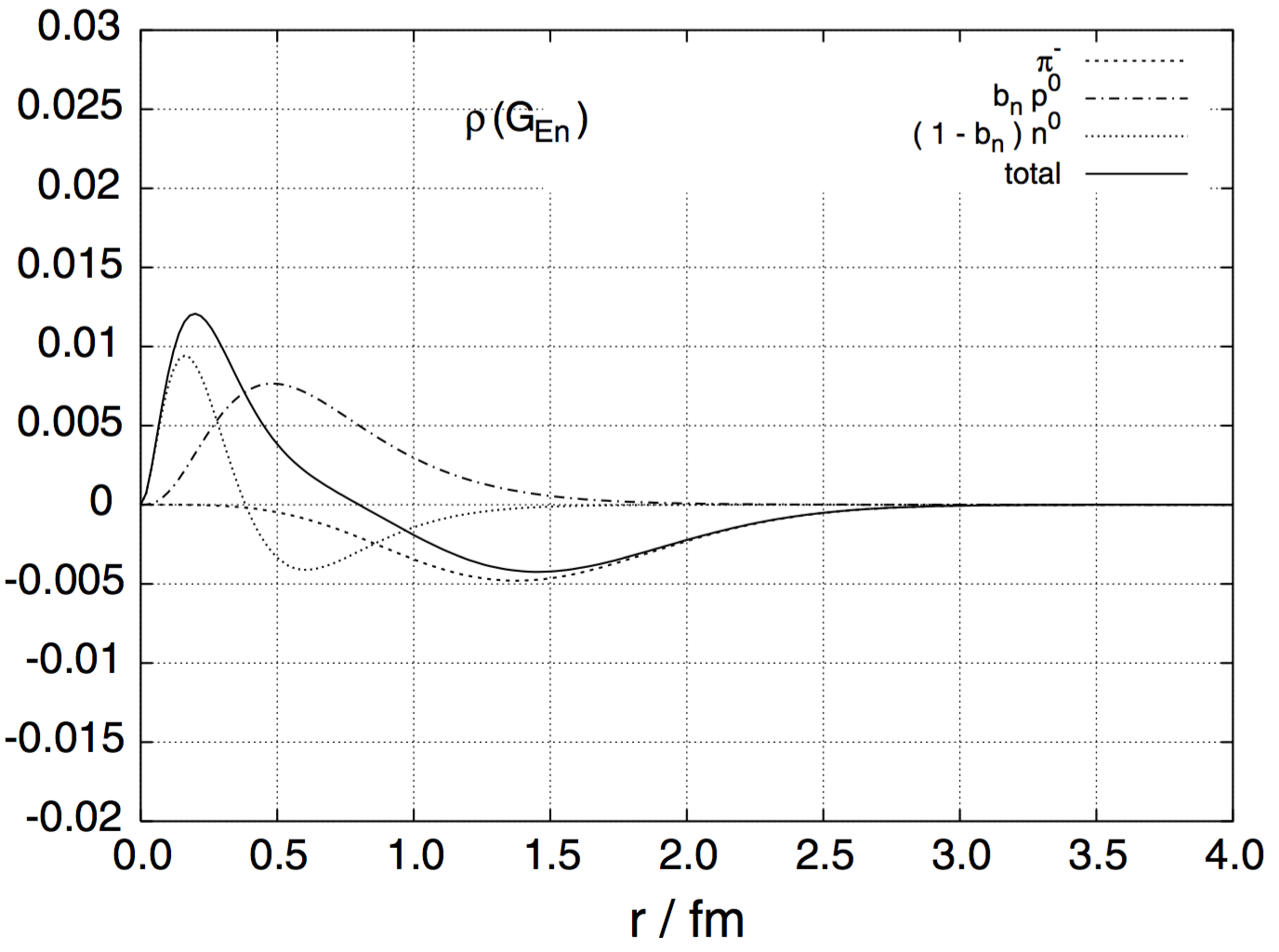}
\end{center}
\caption{\label{fig:GEpGEn} Shown as the dashed lines are the densities of nucleon charge contributed by the pion cloud, taken from Ref.~\cite{Friedrich:2003}. Left panel: $r^2\rho(r)$ for the electric form factor of the proton. Right panel: $r^2\rho(r)$ for the electric form factor of the neutron.}
\end{figure}

\begin{figure}[b!]
\begin{center}
\includegraphics[width=1\textwidth]{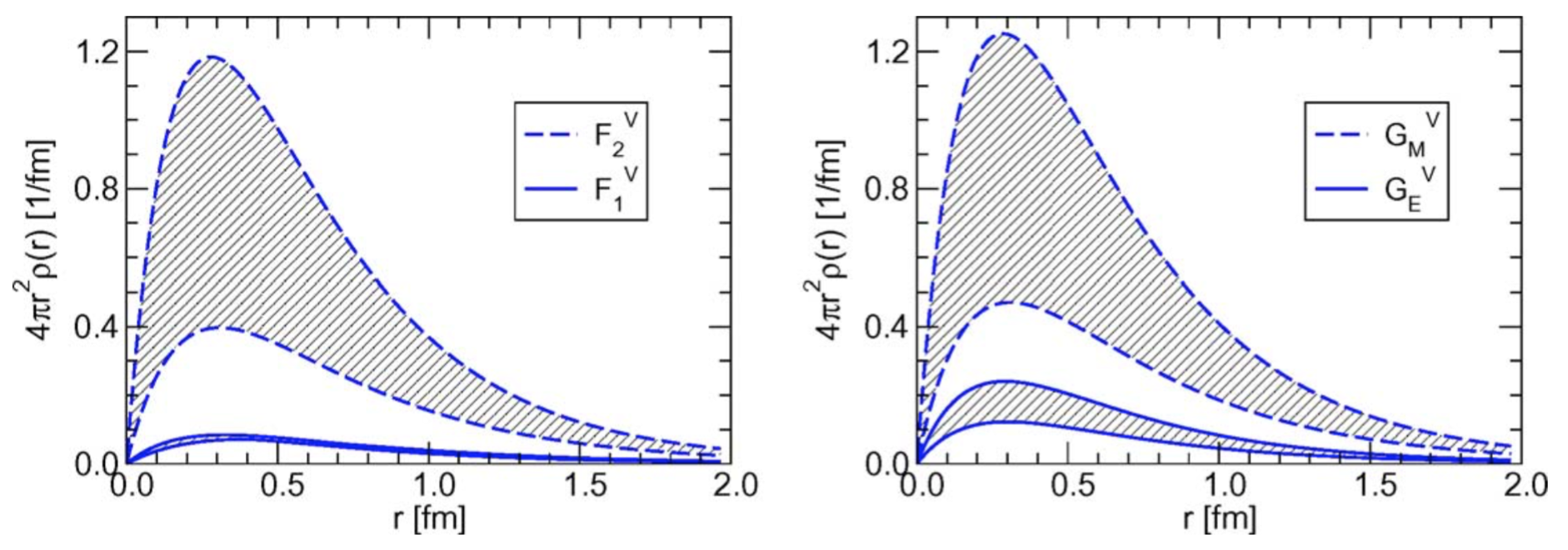}
\end{center}
\caption{\label{fig:GEM} The densities of charge and magnetization from the pion cloud taken from Ref.~\cite{Hammer:2004}. Left panel: $4\pi r^2\rho(r)$ for the isovector Pauli (upper band) and Dirac (lower band) form factors. Right panel: $4\pi r^2\rho(r)$ for the isovector magnetic (upper band) and electric (lower band) Sachs form factors. }
\end{figure}

The investigation of the size or length scale of the meson cloud distribution inside the nucleon is interesting and important since it may help us to understand the in internal structure of nucleon intuitively. In Ref.~\cite{Friedrich:2003}, the meson cloud distribution has been extracted from the nucleon EM form factors in the constituent quark model. The $\pi$-meson cloud distribution is found very long-ranged, $\sim 2$ fm (see Fig.~\ref{fig:GEpGEn} dashed curve), and is interpreted as the result of a pion cloud around the bare nucleon. In comparison with Ref.~\cite{Friedrich:2003}, however, a much more confined $\pi$-meson cloud contribution to the nucleon EM form factors in $r$-space has been derived in the chiral perturbation theory~\cite{Hammer:2004,MeiBner:2007}.  The results in Refs.~\cite{Hammer:2004,MeiBner:2007} reveal that the $\pi$-meson cloud distributions peak around $r=0.3$ fm and fall off smoothly with increasing the distance as shown in Fig.~\ref{fig:GEM}. Similar results have been also obtained in the chiral soliton model~\cite{MeiBner:1988,Holzwarth:1996}. The results in ~\cite{Hammer:2004,MeiBner:2007,MeiBner:1988,Holzwarth:1996} may indicate that there is no structure at larger distances. In this work, we attempt to quantitatively study and define the $r$-space meson cloud distribution inside the nucleon in the framework of the PCQM.

The paper is organized as follows. In section~\ref{sec:PCQM}, we briefly describe the basic notions of the PCQM. The comparison and discussion between EM and axial form factors of nucleon are given in section~\ref{sec:dis}.

\section{\label{sec:PCQM} Perturbative chiral quark model}

In the framework of the PCQM, baryons are considered as the bound states of three relativistic valence quarks moving in a central Dirac field with $V_{\textrm{eff}}(r) = S(r) + \gamma^0 V (r)$, while a cloud of pseudoscalar mesons, as the sea-quark excitations, is introduced for chiral symmetry requirements, and the interactions between quarks and mesons are achieved by the nonlinear $\sigma$ model in the PCQM. The Weinberg-type Lagrangian of the PCQM under an unitary chiral rotation~\cite{Lyubovitskij1:2001,Khosonthongkee:2004} is derived as,
\begin{eqnarray}
\mathcal{L}^W(x)&=& \mathcal{L}_0(x)+\mathcal{L}^W_I(x)+o(\vec{\pi}),\label{eq:WT-Lagrangian}\\
\mathcal{L}_0(x)&=& \bar{\psi}(x)\big[i\partial\!\!\!/-\gamma^{0}V(r)-S(r)\big]\psi(x)
-\frac{1}{2}\Phi_i(x)\big(\Box+M_{\Phi}^2 \big) \Phi^i(x),\label{WT-0}\\
\mathcal{L}_I^W(x)&=&\frac{1}{2F}\partial_\mu\Phi_i(x)\bar{\psi}(x)\gamma^\mu\gamma^5
\lambda^i\psi(x)+\frac{f_{ijk}}{4F^2}\Phi_i(x)\partial_\mu\Phi_j(x)
\bar\psi(x)\gamma^\mu\lambda_k\psi(x)\label{eq:WT-int},
\end{eqnarray}
where $f_{ijk}$ are the totally antisymmetric structure constant of $SU(3)$, the pion decay constant $F=88$ MeV in the chiral limit, $\Phi_i$ are the octet meson fields,
and $\psi(x)$ is the triplet of the $u$, $d$, and $s$ quark fields taking the form
\begin{equation}
\psi(x) =\left(
\begin{array}{c}
u(x) \\
d(x) \\
s(x) \\
\end{array}
\right).
\end{equation}
The quark field $\psi(x)$ could be expanded in
\begin{equation}
\psi(x)=\sum_\alpha\left(b_\alpha u_\alpha(\vec{x})\,e^{-i\mathcal{E}_\alpha t}+d^\dagger_\alpha\upsilon_\alpha (\vec{x})e^{i\mathcal{E}_\alpha t}\right),
\end{equation}
where $b_\alpha$ and $d^\dag_\alpha$ are the single quark
annihilation and antiquark creation operators. The ground state quark wave function $u_0(\vec x)$ may, in general, be expressed as
\begin{equation}
u_0(\vec{x})=\left(\begin{array}{c}g(r)\\\large{i\vec{\sigma}\cdot\hat{x}f(r)}
\end{array}\right)\chi_s\chi_f\chi_c,
\end{equation}
where $\chi_s$, $\chi_f$ and $\chi_c$ are the spin, flavor and color quark wave functions, respectively.

\begin{figure}[h!]
\begin{center}
\includegraphics[width=0.31\textwidth]{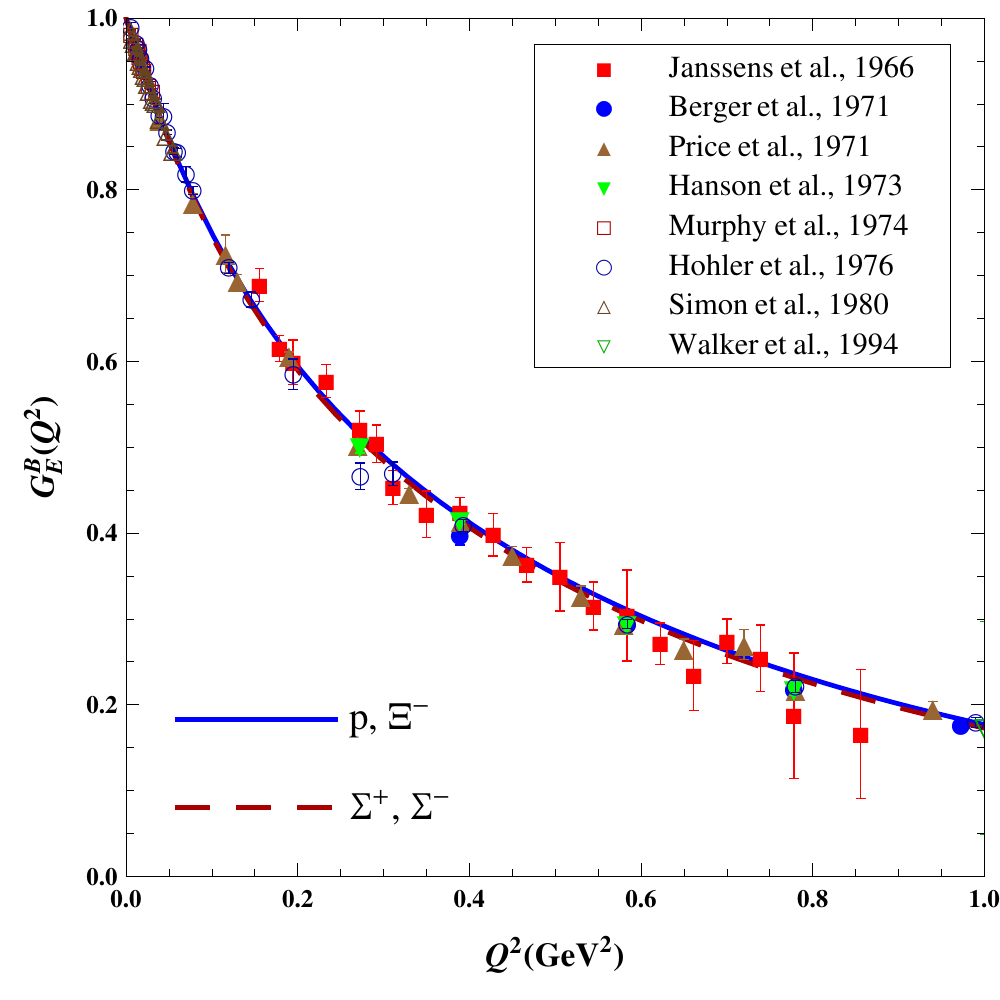}
\hspace{0.4cm}
\includegraphics[width=0.31\textwidth]{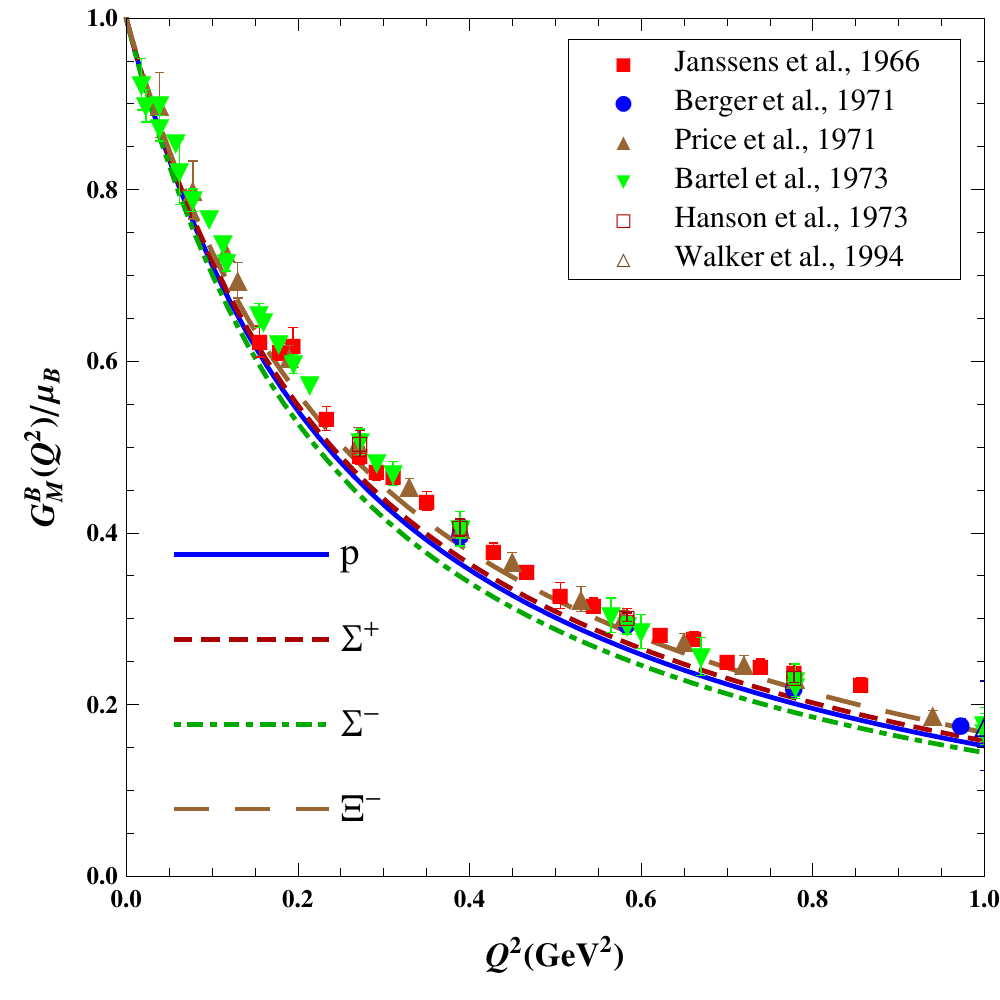}
\hspace{0.4cm}
\includegraphics[width=0.31\textwidth]{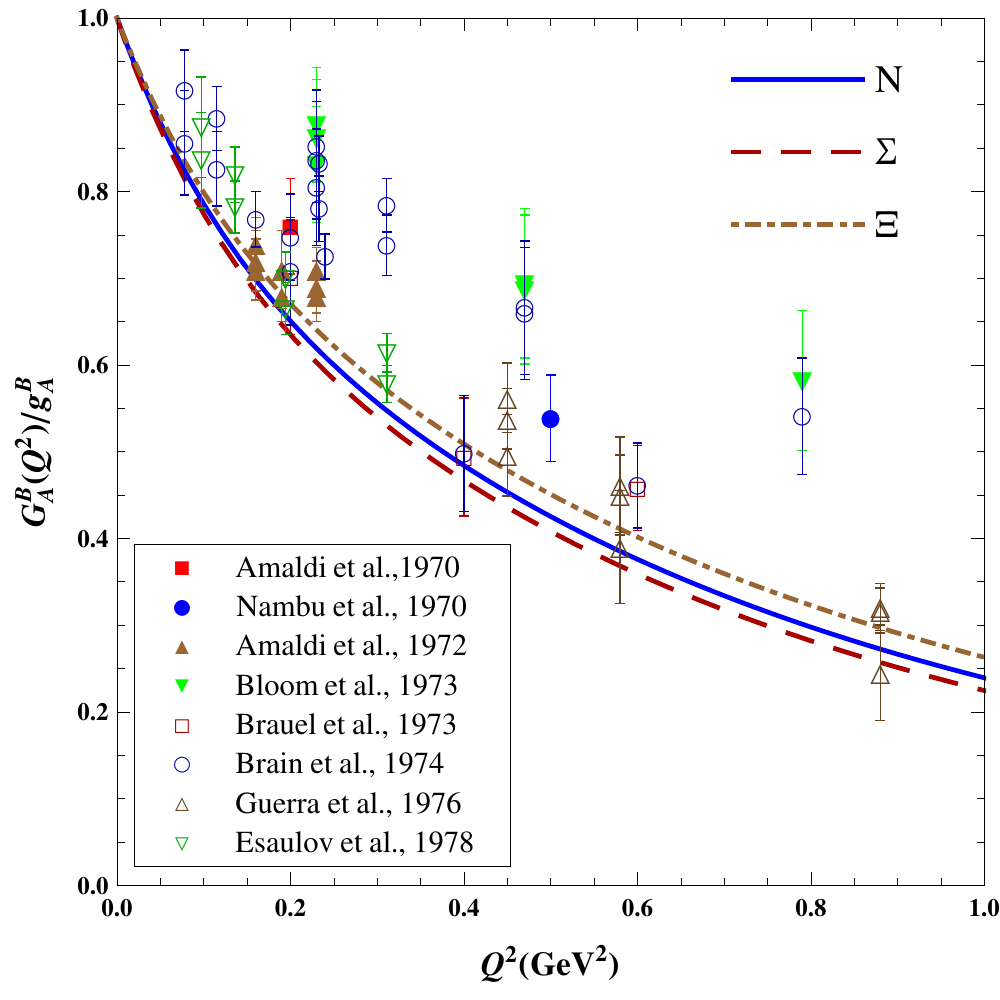}
\end{center}
\caption{\label{fig:GEMB} The results of EM and axial form factors in the PCQM taken from Refs.~\cite{Liu:2014,Liu:2015}.}
\end{figure}

In our previous works~\cite{Liu:2014,Liu:2015}, the ground state quark wave functions have been determined by fitting the PCQM theoretical result of the proton charge form factor $G_E^p(Q^2)$ to the experimental data~\cite{Liu:2014}, and the electromagnetic and axial form factors as well as electroweak properties of octet baryons in low energy region have been studied in the PCQM based on the predetermined quark wave functions. As the results shown in Fig.~\ref{fig:GEMB}, the EM and axial form factors are in good agreement with the experimental data. Meanwhile, the nucleon magnetic moment $\mu_p=2.735\pm0.121$ and axial charge $g_A^N=1.301\pm0.230$, which are the magnetic and axial form factors in zero-recoil, are consistent with the experimental data and the lattice QCD values. One may indicate that the PCQM is credible, and able to further discuss the meson cloud distributions. More details could be found in Refs.~\cite{Liu:2014,Liu:2015}.

\section{\label{sec:dis} Electric and axial charge distributions of meson cloud}

\begin{figure}[t!]
\begin{center}
\includegraphics[width=0.45\textwidth]{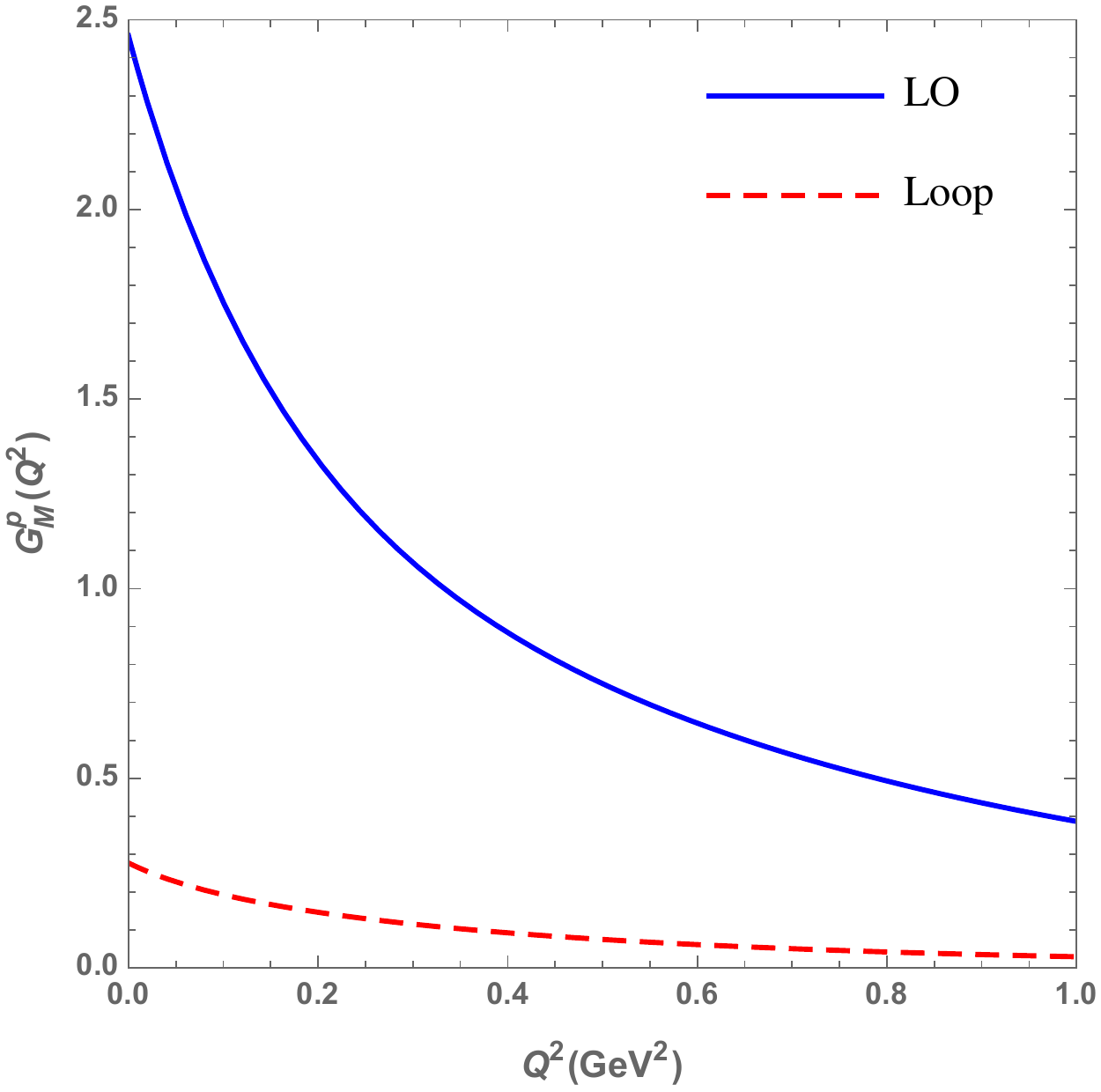}
\hspace{0.5cm}
\includegraphics[width=0.45\textwidth]{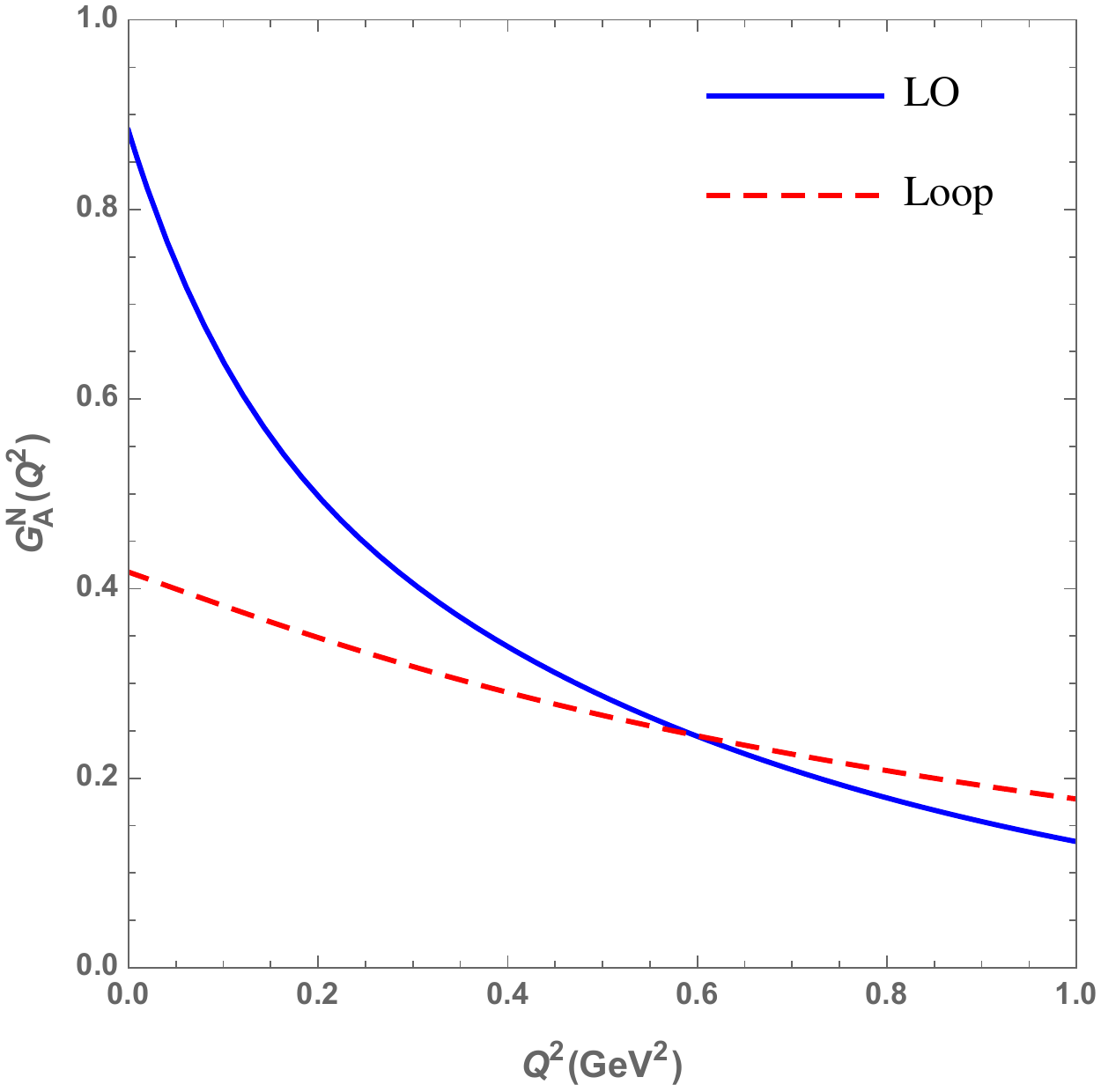}
\end{center}
\caption{\label{fig:GEMAN} Leading order (solid) and loop (dashed) contributions to the proton magnetic (left panel) form factor and neutron axial (right panel) form factor.}
\end{figure}

Following our previous works~\cite{Liu:2014,Liu:2015}, we present in Fig.~\ref{fig:GEMAN} the proton magnetic and nucleon axial form factors separately in leading order (LO) and loop Feymann diagram contributions.  The PCQM results shown in Fig.~\ref{fig:GEMAN} clearly reveal that the LO diagram results in a dipole-like form factor while the meson cloud leads to a flat contribution to the magnetic and axial form factors. The flat contribution may indicate that the meson cloud of the nucleon may distribute mainly in a very small region.

\begin{figure}[b!]
\begin{center}
\includegraphics[width=0.45\textwidth]{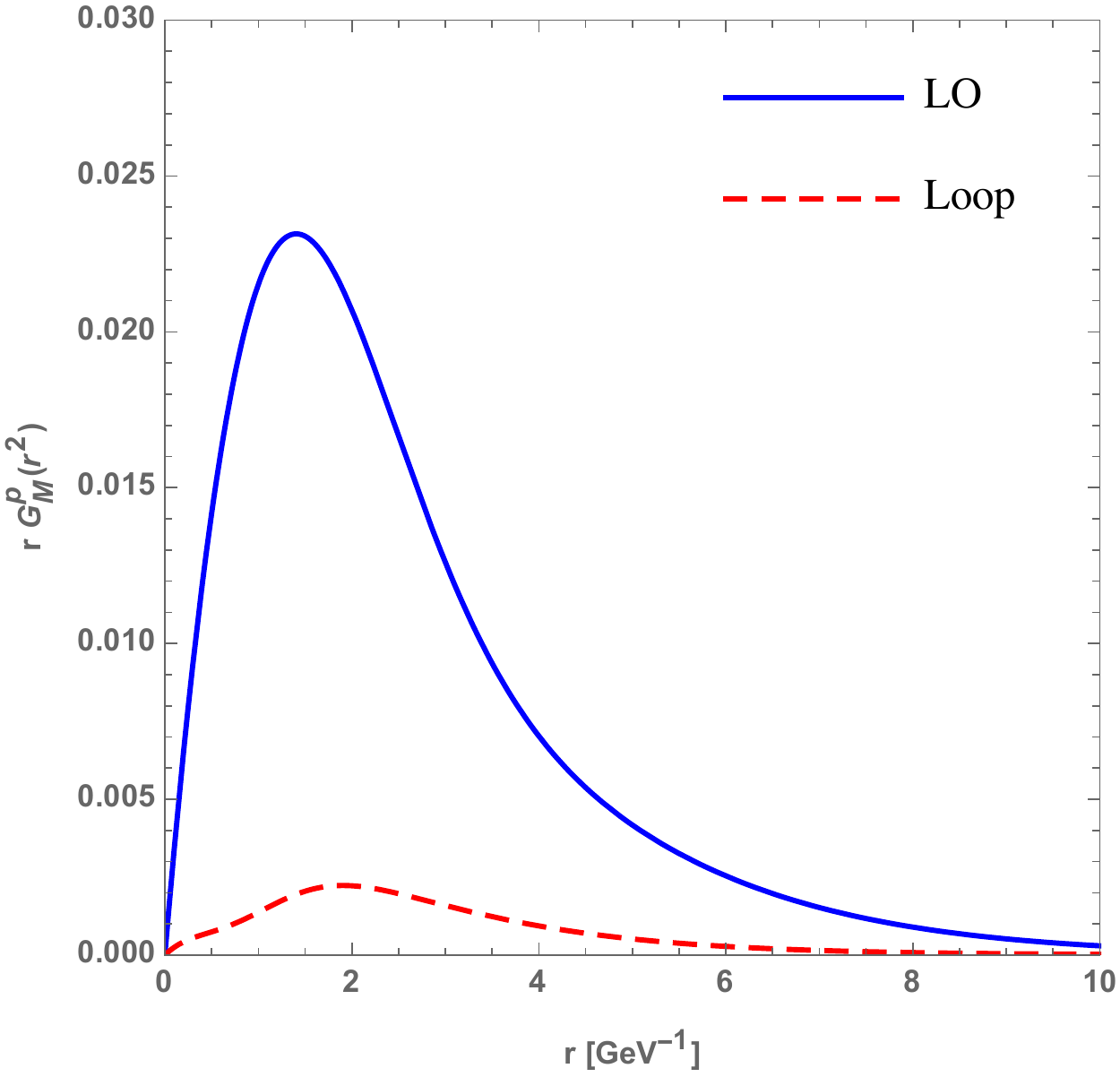}
\hspace{0.5cm}
\includegraphics[width=0.45\textwidth]{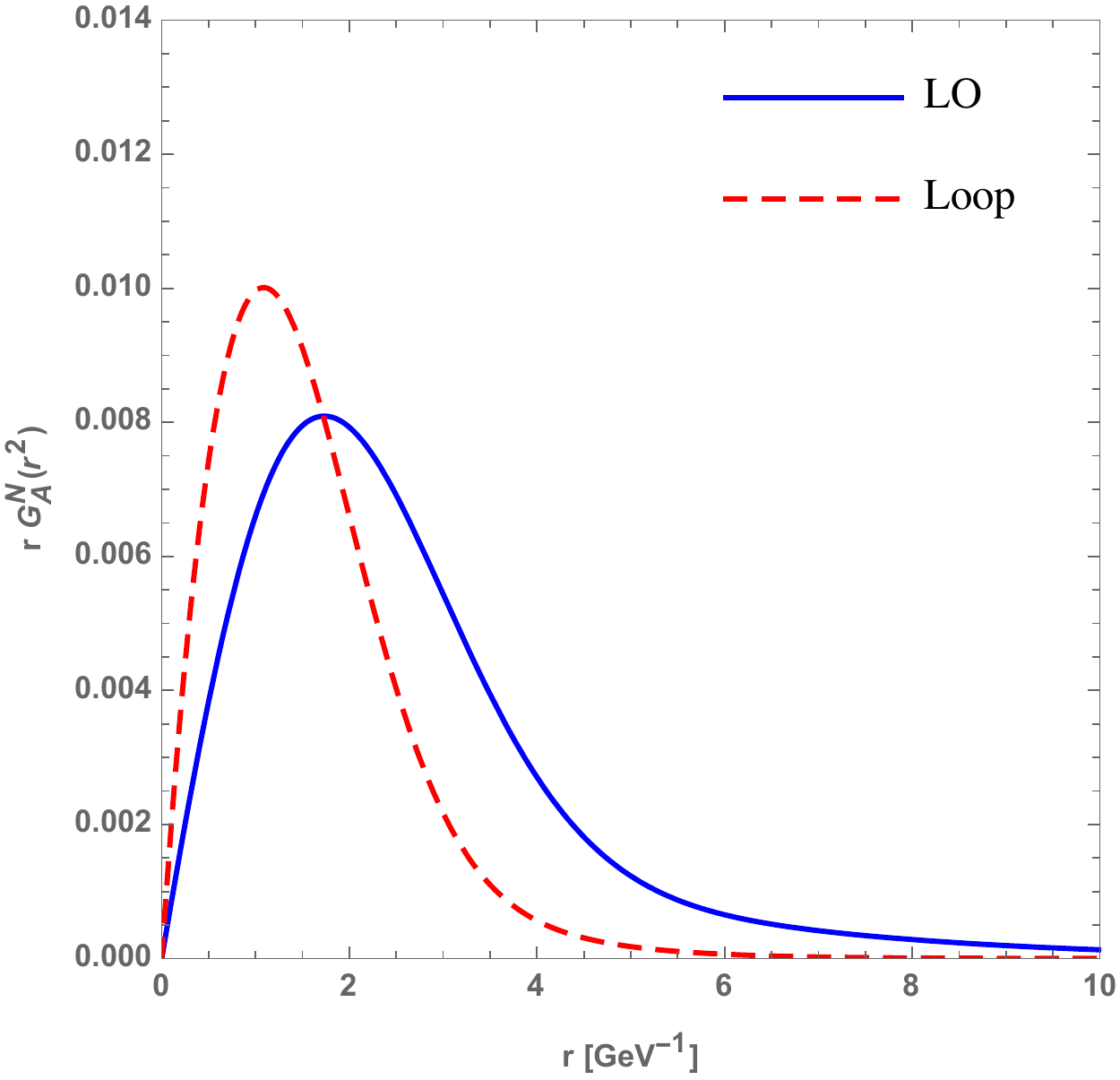}
\end{center}
\caption{\label{fig:LOLoop} Comparisons between the LO and meson cloud distributions for proton magnetic (left panel) and axial (right panel) form factors in $r$-space.}
\end{figure}

\begin{figure}[t!]
\begin{center}
\includegraphics[width=0.45\textwidth]{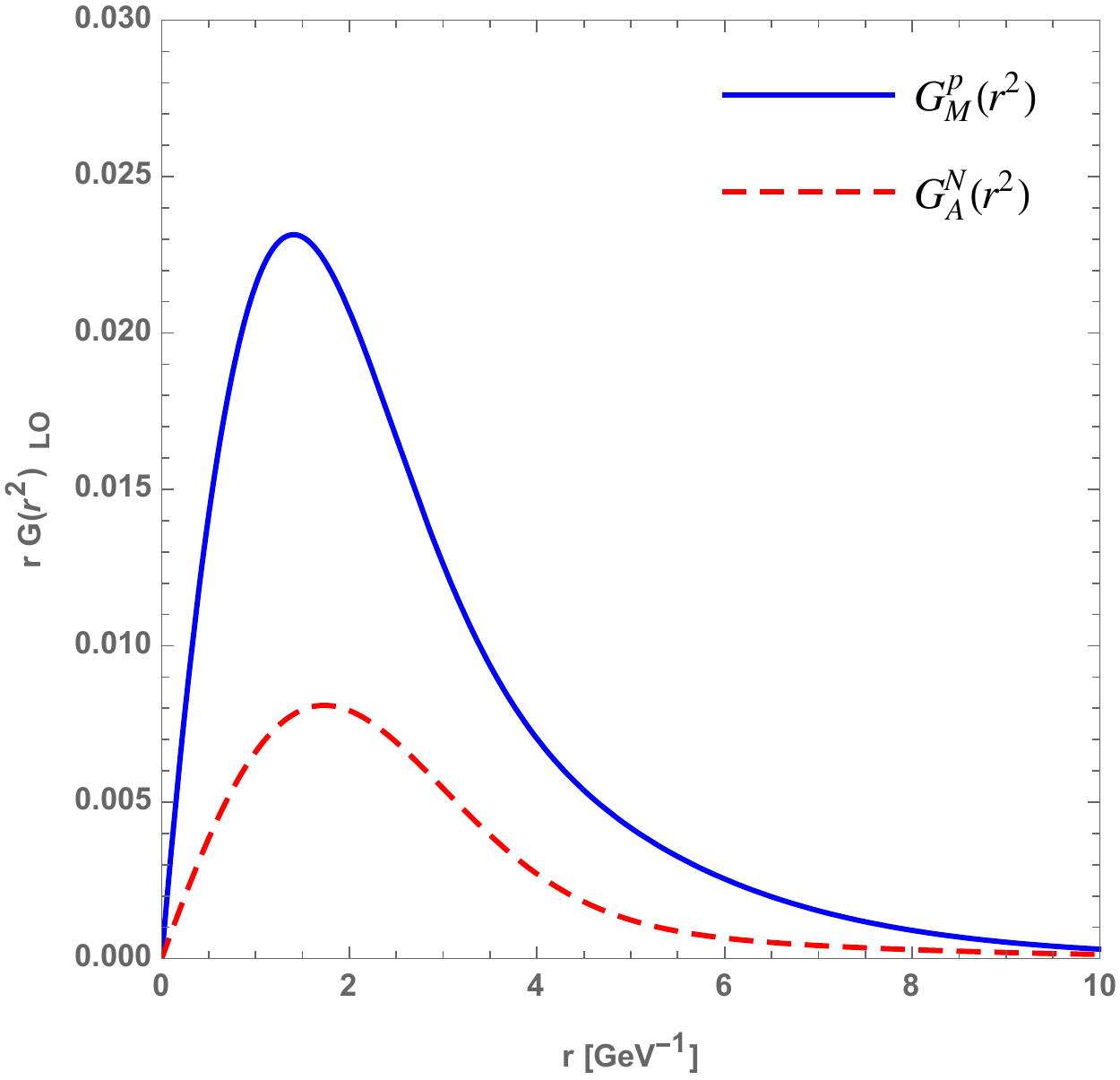}
\hspace{0.5cm}
\includegraphics[width=0.455\textwidth]{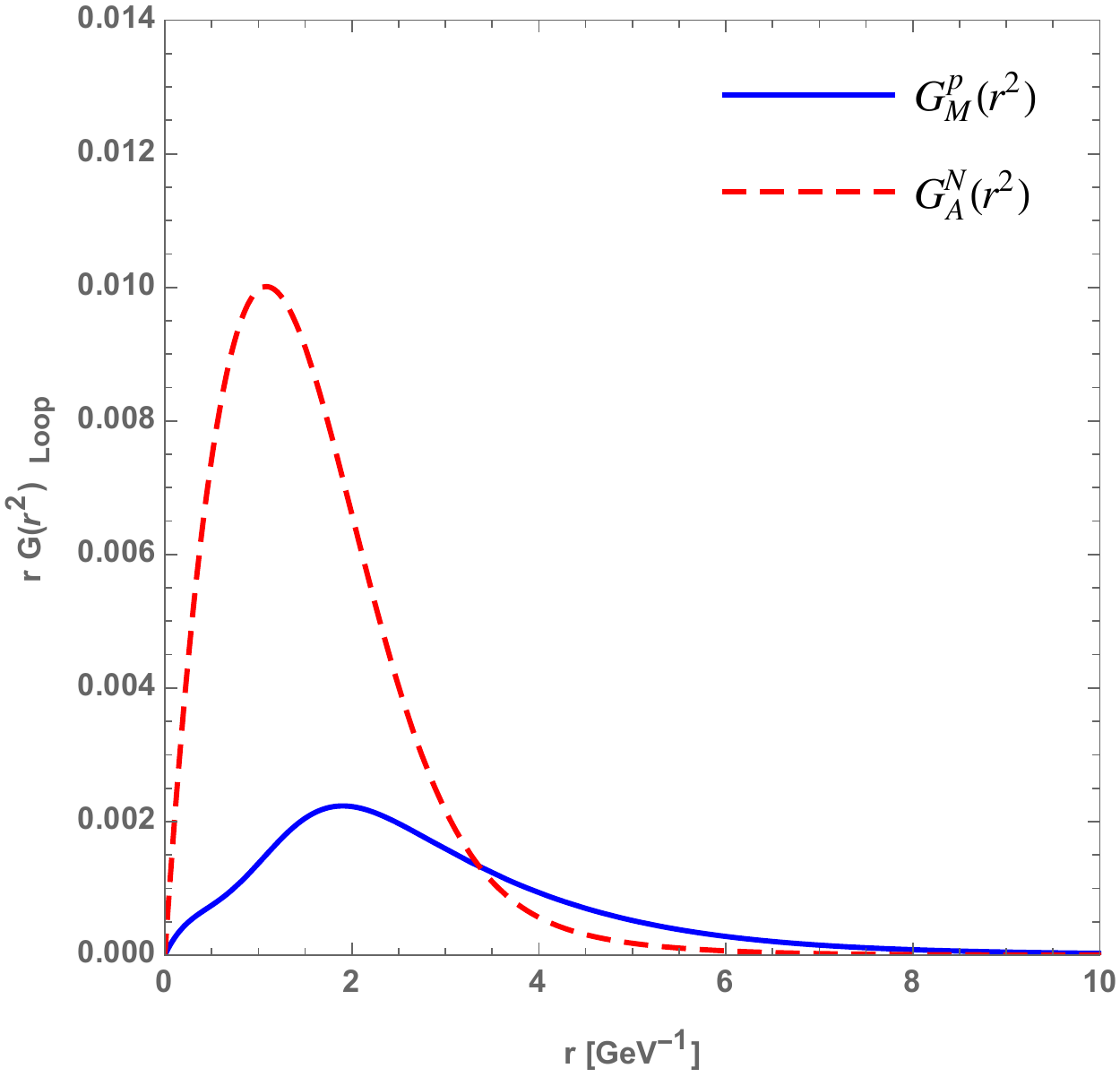}
\end{center}
\caption{\label{fig:GMpGAN} Comparisons between the magnetic and axial distributions in $r$-space for the LO (left panel) and loop (right panel) diagrams.}
\end{figure}

In this work, we extract, based on the inverse Fourier transformation, the magnetic charge and axial charge meson cloud distributions of the nucleon from the EM and axial form factors as shown in Fig.~\ref{fig:GEMAN}. Shown in Fig.~\ref{fig:LOLoop} are the LO and meson cloud contributions to the proton magnetic form factor $G_M^p(Q^2)$ (left panel) and the nucleon axial form factor $G_A^N(Q^2)$ (right panel) in $r$-space.  It is found in Fig.~\ref{fig:LOLoop} that the magnetic charge distributions of the three-quark core and the meson cloud are almost in the same region, but the loop diagrams contribute to the $G_A^N(Q^2)$ in a clearly smaller region than the LO diagram, which may indicate that the axial charge meson cloud distributes mainly inside the three-quark core.

We also find that the magnetic charge distribution shown in the left panel of Fig.~\ref{fig:LOLoop} present a significant peak around $r\simeq2$ $\rm GeV^{-1}$ and fall off smoothly when the distance increases. The results turned out to be similar to the ones of Refs.~\cite{Hammer:2004,MeiBner:2007}.

Furthermore, we compare the LO contributions to the $G_M^p(Q^2)$ and $G_A^N(Q^2)$ in $r$-space, as presented in the left panel of Fig.~\ref{fig:GMpGAN}. It is clear that the LO $G_M^p(r^2)$ and $G_A^N(r^2)$ show a similar $r$-dependence, which may indicate that the electric charge and axial charge distributions of the constituent quarks are the same. The meson cloud $G_M^p(r^2)$ and $G_A^N(r^2)$ in the right panel of Fig.~\ref{fig:GMpGAN} show that the axial charge distribution of the meson cloud is narrower and the peak is closer to the origin.

In summary, one may conclude that the similar r-dependence of the magnetic and axial form factors resulted from the LO diagrams may indicate that the electric charge and axial charge distributions of the constituent quarks are the same. The magnetic charge distributions of the meson cloud and three-quark core are more or less in the same region and peak at distances of around 2 $\rm GeV^{-1}$, quite consistent with earlier determinations in Refs.~\cite{Hammer:2004,MeiBner:2007}, but the axial charge meson cloud distributes mainly inside the three-quark core.

\section*{Acknowledgments}

This work is supported by National Natural Science Foundation of China (Project No. 11547182), and the Doctoral Scientific Research Foundation of Liaoning Province (Project No. 201501197). This work is also supported by Suranaree University of Technology and Bohai university. XL and AL acknowledge support by SUT-CHE-NRU (Project No. NV12/2558). 

\section*{References}

\bibliographystyle{elsarticle-num}
\bibliography{Refs}

\end{document}